\documentclass[preprint,showpacs,preprintnumbers,amsmath,amssymb]{revtex4}
\usepackage{graphicx}
\usepackage{dcolumn}
\usepackage{bm}
\usepackage{hyperref}

\begin{document}

\title{Unstable states in dissociation of relativistic nuclei.\\ Recent findings and prospects of researches.}

\author{D.A. Artemenkov$^1$}
\author{V. Bradnova$^1$}
\author{M.M. Chernyavsky$^2$}
\author{E. Firu$^3$}
\author{M. Haiduc$^3$}
\author{N.K. Kornegrutsa$^1$}
\author{A.I. Malakhov$^1$}
\author{E. Mitsova$^1$}
\author{A. Neagu$^3$}
\author{N.G. Peresadko$^2$}
\author{V.V. Rusakova$^1$}
\author{R. Stanoeva$^4$}
\author{A.A. Zaitsev$^{1,2}$}
\author{I.G. Zarubina$^1$}
\author{P.I. Zarubin$^{1,2}$}
\email[e-mail:]{zarubin@lhe.jinr.ru}

\affiliation {${}^{1}$Joint Institute for Nuclear Research, Dubna, Russia \\
			  ${}^{2}$Lebedev Physical Institute, Russian Academy of Sciences, Moscow, Russia \\
			  ${}^{3}$Institute of Space Science, Magurele, Romania\\
			  ${}^{4}$Southwestern University, Blagoevgrad, Bulgaria }

\begin{abstract}
Results are presented on the identification of the unstable nuclei $^{8}$Be and $^{9}$B and the Hoyle state (HS) in the relativistic dissociation of the isotopes $^{9}$Be, $^{10}$B, $^{10}$C, $^{11}$C, $^{12}$C, and $^{16}$O in a nuclear track emulsion (NTE). The main motivation for the study is the prospect of using these states in the search for more complex unstable states that decay with their participation. The possibilities of the NTE method for studying the contribution of multiple ensembles of the lightest He and H nuclei to the fragmentation of relativistic nuclei are described in brief. It is shown that to identify relativistic decays $^{8}$Be and $^{9}$B and HS in NTE, it is sufficient to determine the invariant mass as a function of angles in pairs and triples of He and H fragments in the approximation of conservation of momentum per nucleon of the parent nucleus. The formation of HS in the dissociation $^{16}$O $\to$ 4$\alpha$ is observed. According to the criteria established in this way, the contribution of the unstable states to the relativistic fragmentation of $^{28}$Si and $^{197}$Au nuclei was estimated. Promising applications of the NTE method in the study of nuclear fragmentation are discussed.
\end{abstract}


\maketitle

\section{Introduction}
\label{intro}
The generation of ensembles consisting of several He and H nuclei is possible in the peripheral dissociation of relativistic nuclei. Potentially, an in-depth study of their features can shed light on topical issues in few-body nuclear physics. The focus of theoretical developments is the possibility of the existence of states with a pronounced $\alpha$-condensate and nuclear-molecular structure. In turn, the findings of the corresponding laboratory searches could be involved in the development of multi-body scenarios of nuclear astrophysics.

Being flexible and inexpensive, the method of nuclear track emulsion (NTE) meets the corresponding experimental challenges at least at the search stage. In the NTE layers longitudinally exposed to relativistic nuclei tracks of fragments can be observed exhaustively, and their direction determined with the best resolution. Determination of the invariant mass of the relativistic fragment groups in the approximation of conservation of the velocity of the initial nucleus makes it possible to project the angular correlations of fragments on the energy scale of nuclear physics. Thus, on the basis of the relativistic-invariant approach, a new and at the same time visual paradigm of the experimental study of ensembles of the lightest nuclei just above the coupling threshold appears. It is possible both to use the results obtained within the framework of the NTE method when planning experiments of high complexity and a variety of the most modern detectors, as well as to apply in a large-scale the method itself leaning on the achievements of computerized microscopy. In this context, a brief overview of the application of NTE to unstable $^{8}$Be and $^{9}$B nuclei and the search for heavier states associated with them will be given below.

Technological and analytical foundations of the NTE method as applied to relativistic particles were developed in the initial period when fundamental discoveries in high-energy physics were made in the study of cosmic rays. These achievements are fully presented in the classic book by C.H. Powell, P.H. Fowler and D.H. Perkins \cite{1} along with photographs of characteristic events. Its last chapter is devoted to discovery of relativistic nuclei in cosmic radiation.

Tracks of relativistic nuclei of cosmic origin and nuclear stars created by them were found in NTE stacks irradiated in stratosphere at the end of the 40s \cite{2}. To describe the cross section for interaction of these nuclei with nuclei of the NTE composition, a geometric overlap formula was proposed, later called the Bradt-Peters formula. Among others, stars were observed containing few relativistic $\alpha$ particle tracks near directions of parent nucleus tracks. Defined as peripheral, such interactions clearly indicated the $\alpha$-partial clustering of the final state. This aspect of the nuclear structure is studied in depth in low energy nuclear reactions by precision spectroscopic methods.

In the 70s, exposures of NTE stacks to few GeV per nucleon light nuclei started at the JINR Synchrophasotron and Bevalac LBL, and in the 90s, medium and heavy ones on AGS (BNL) and SPS (CERN) at significantly higher energy values. With a resolution of about \mbox{0.5 $\mu$m} the NTE method provided exclusive observations of tracks from the heaviest relativistic nuclei down to singly charged particles and slow fragments of target nuclei. A search for stars along primary tracks made it possible to make an overview of types of interaction without sampling. The results obtained from 70s to the present by this method, as well as the corresponding data files, remain unique in terms of the composition of relativistic fragmentation, and the exposed NTE layers can be used for targeted studies.

Features of light nuclei appeared in the relativistic fragmentation cone (for example, \cite{3}). These include the universal formation of pairs of $\alpha$ particles with extremely small opening angles, of the order of few mrad. Such narrow ``forks'' correspond to decays of the unstable $^{8}$Be nucleus. They indicate the principal possibility of studying $\alpha$-clustering in relativistic approach starting from the $^{8}$Be decay energy.

Until now, the complete detection of ensembles of the lightest relativistic fragments has been demonstrated only by the NTE method. However, it does not provide the momentum analysis. This limitation can be compensated for by using information on the fragmentation of relativistic nuclei obtained from magnetic spectrometers (for example, \cite{4,5}). In the context of this article, it is worth noting the availability of data obtained in 90s on an exclusive study of the fragmentation of relativistic oxygen nuclei on protons using the JINR hydrogen bubble chamber placed in a magnetic field \cite{6}.

Experiments in beams of fast-moving nuclei devoted to the nuclear structure provide both advantages of detection and make it possible to include radioactive isotopes, including exotic ones, among the studied ones. Information about the structure of ground states and peripheral interactions of nuclei is extracted from momentum distributions of relativistic fragments with charges close to the initial nucleus (for example, \cite{7,8,9}). The formation of the relativistic $^{7}$Be$^*$ fragment ($^{8}$B $\to$ $^{7}$Be + $\gamma$) \cite{10} in an only bound excitation has been proved \cite{11}. Investigations of nuclear excitations with registration together with a fragment of one or a pair of protons or neutrons require a transition to the region of hundreds of MeV per nucleon (for example, \cite{12} and recent \cite{13} ). Then the primary beam and charged fragments, which are quite different in magnetic rigidity, can be separated.

However, in the case of light nuclei, the role of channels with only He and H fragments is a key one, at least due to the fact that the $^{8}$Be and $^{9}$B isotopes whose presence in the cone of fragmentation is very likely are unstable ones. For example, according to the NTE data $^{11}$C dissociation channels containing only He and H make up 80\% of events with the primary charge remaining in the fragmentation cone. Being part of the fragmentation pattern reconstruction of $^{8}$Be and $^{9}$B is necessary for the subsequent searches of unstable states decaying with their participation.

Collimation of relativistic fragments produced in a peripheral collision allows detecting all of them all at once in a small solid angle that is an obvious value. However, electronic experiments in this direction have encountered difficulties due to the quadratic dependence of ionization on charges, the extremely small angular divergence of relativistic fragments, and, often, an approximate coincidence in magnetic rigidity with the beam nuclei. For instance, in the experiment devoted to dissociation $^{12}$C $\to$ 3$\alpha$ at 2.1 $A$ GeV, it was not possible to identify $^{8}$Be decays and the bulk of the energy spectrum of $\alpha$ triples was reconstructed by simulation \cite{14}. It seems that the main problem is that an increase in the degree of dissociation of the projectile leads to a rapid decrease in the ionization signal in the detectors and, apparently, unacceptably increasing requirements for their operating range. A magnetic spectrometer with a time-projection chamber with a wide sensitivity range is being developed at the GSI \cite{15}. The experiment is aimed at studying the isotopic composition and fragmentation mechanisms of 1 \textit{A} GeV heavy nuclei. In case of interest in the topic of clustering, the possibility of reconstructing the dissociation $^{12}$C $\to$ $^{8}$Be + $\alpha$ is worth considering.

Since the early 2000s the NTE method was applied in the BECQUEREL experiment at the JINR Nuclotron to study, in the relativistic approach, the composition of light fragmentation of several stable and radioactive nuclei (reviewed in \cite{16,17,18}). For this experiment the Slavich Company (Pereslavl Zalessky, Russia) has resumed production of NTE layers with a thickness from 50 to 200 $\mu$m on a glass base. NTE samples were tested with a whole variety of ionization tracks. At present, the production of layers with a thickness of 500 $\mu$m without a substrate is being mastered, which will allow continuing the application of the technique, which was considered almost lost.

So, known and previously unobserved structural features revealed of the isotopes $^{7,9}$Be, $^{8,10,11}$B, $^{10,11}$C, and $^{12,14}$N are revealed in the dissociation channel probabilities. Decays \mbox{$^{9}$B $\to$ $^{8}$Be + \textit{p}} are identified in the dissociation $^{10}$B, $^{10}$C, and $^{11}$C. Earlier, dozens of $^{9}$B decays were identified in fragmentation of 200–400 \textit{A} MeV $^{12}$C nuclei in a water target, when tracks were reconstructed in transversely placed NTE films \cite{19}.Apparently, the absence of a stable state of the $^{9}$B nucleus does not prevent its virtual presence in the structure of these nuclei. Their synthesis could occur through the $^{9}$B + \textit{p} resonance along the chain $^7$Be($^3$He,\textit{p})$^9$B(\textit{p},$\gamma$)$^{10}$C(e$^+$,$\nu$)$^{10}$B(\textit{p},$\gamma$)$^{11}$C(e$^+$,$\nu$) $^{11}$B. As a result, $^{9}$B is ``imprinted'' in the formed nuclei, which is manifested in relativistic dissociation. In the $^{7}$Be fragmentation, $^{6}$Be $\to$ $\alpha$ + 2\textit{p} decays are identified. However, the $^{6}$Be signal was not detected in the $^{10}$C dissociation.

With further advancement along the border of proton stability, the possibility of effective study of multiple final states He and H remains. At the same time, such an approach is limited by the impossibility of isotopic identification of fragments heavier than He whose contribution rapidly increases with increasing mass number of the nucleus under study. Their identification is possible in experiments with magnetic analysis.

Next, the problems of unstable states and the results of their searches in the relativistic dissociation of several light nuclei in NTE will be summarized. Being interesting with respect to the structure of the studied nuclei these observations allow one to address the question of their universality, including their manifestation in the dissociation of medium and heavy nuclei, where it becomes possible to search for increasingly complex unstable states. As a first step, an analysis of NTE exposed in BNL to 14.5 \textit{A} GeV $^{28}$Si and 10.7 \textit{A} GeV $^{197}$Au nuclei is presented.

\section{Problems of unstable states}
\label{sec:1}

The identification of the relativistic $^{8}$Be and $^{9}$B decays in NTE pointed out the possibility of identifying the unstable state of the triple of $\alpha$ particles, called the Hoyle state (HS) in the relativistic dissociation $^{12}$C $\to$ 3$\alpha$ \cite{20} and, then, $^{16}$O $\to$ 4$\alpha$ \cite{21}. The solution to this problem allows HS to be used as a ``tool'' for searching for exotic components in the nuclear structure and complex unstable states decaying with its participation. It is worth studying the possibility of extracting information about the size of HS based on the distributions of the total transverse momentum of $\alpha$-triples.

HS is the second (and first $\alpha$-unbound) 0$^+_2$ excitation of the $^{12}$C nucleus \cite{11}. The discovery history and research status of this short-lived state of three real $\alpha$ particles are discussed in a review \cite{22}. $^{12}$C synthesis is possible through two unstable states 3$\alpha$ $\to$ $\alpha$$^{8}$Be $\to$ \mbox{$^{12}$C (0$^+_2$ or HS)} $\to$ $^{12}$C. In the 3$\alpha$-process, HS manifests itself as an unstable nucleus, albeit of an unusual nuclear molecular structure.

The nucleus $^{8}$Be is an indispensable product of the decay of HS and $^{9}$B. The decay energy of $^{8}$Be is 91.8 keV, and the width is 5.57 $\pm$ 0.25 eV \cite{11}. The isolation of HS among $^{12}$C excitations, the extremely small values of the energy above the 3$\alpha$ threshold (378 keV) and the decay width (9.3 $\pm$ 0.9 eV) indicate similarity with the $^{8}$Be nucleus \cite{11}. The $^{9}$B ground state is 185.1 keV higher than the threshold $^{8}$Be + \textit{p} and its width is 0.54 $\pm$ 0.21 keV \cite{11}. A comparison of these parameters suggests that the significance of HS to nuclear physics is not limited to the role of the unusual excitation of the $^{12}$C nucleus. HS is manifested in nuclear reactions as the universal object similar to $^{8}$Be and $^{9}$B \cite{23,24,25}.

According to their widths, $^{8}$Be, $^{9}$B, and HS can be full participants of relativistic fragmentation. The products of their decay are formed during runs from several thousand ($^{8}$Be and HS) to several tens ($^{9}$B) of atomic sizes, i.e., over a time many orders of magnitude longer than the time of the appearance of other fragments. Due to the lowest decay energy, $^{8}$Be, $^{9}$B, and HS should manifest themselves as pairs and triples of relativistic fragments of He and H with the smallest opening angles which distinguishes the latter from other fragmentation products.

$^{8}$Be and HS are considered as the simplest states of the $\alpha$-particle Bose – Einstein condensate \cite{26,27}. The 6$^{th}$ excited state 0$^+_6$ of the $^{16}$O nucleus at 15.1 MeV (or 660 keV over the 4$\alpha$ threshold) is considered as a 4$\alpha$-condensate. Its decay could go in the sequence $^{16}$O(0$^+_6$) $\to$ $^{12}$C(0$^+_2$) $\to$ $^{8}$Be(0$^+_2$) $\to$ 2$\alpha$. Research in this direction is actively underway \cite{23,24,25}. However, the contribution of 4$\alpha$ ensembles above 1 MeV is dominant. The possibility of more complex $\alpha$-condensate states up to 10$\alpha$-particle one with the decay energy of about 4.5 MeV above the 10$\alpha$-threshold is assumed which leads to unprecedented experimental requirements including parent nucleus energy growth.

In addition, the $^{9}$B and HS can serve as bases in the nuclear molecules $^{9}$B\textit{p}, $^{9}$B$\alpha$, and HS\textit{p}. Like $\alpha$-condensate states, unstable states involving protons can correspond to excitations having electromagnetic decay widths. The ratio of the probabilities of $^{13}$N decays on the $^{9}$B$\alpha$ and HS\textit{p} channels is of interest. Excitation of the $^{13}$N$^*$ at 15.1 MeV having a width of 0.86 $\pm$ 0.12 keV \cite{11} can serve as a candidate for this state. An effective source for such studies is the $^{14}$N nucleus. In this regard, the analysis of $^{14}$N dissociation in nuclear reactors via the 3He + H channel is resumed.

Several isotopes have excited states with widths of the order of few eV or lifetimes of several fsec located above the separation thresholds no higher than about 1 MeV of the $\alpha$ particle and the stable residue heavier than He \cite{11}. When such states are formed in the fragmentation, the decay products will also have minimal opening angles. They will be an even more convenient subject of research than $\alpha$\textit{p} states. In this regard, it is planned to analyze the mirror channels $^{11}$C ($^{11}$B) $\to$ $^{7}$Be ($^{7}$Li) + $\alpha$. There is NTE for such an analysis for such an analysis for $^{10}$B, $^{16}$O, $^{22}$Ne, $^{24}$Mg, $^{28}$Si nuclei.

On the whole, the topic of studying unstable states seems unusually intriguing, and the NTE method is an adequate way to search for them in the peripheral interactions of relativistic nuclei. Questions may be raised about the contribution to the fragmentation of decays from more highly excited states with widths up to 100 keV which would also have ranges significantly exceeding the characteristic sizes of nuclei. However, the answers are outside the resolution of the NTE method.

\section{Capabilities of NTE method}
\label{sec:2}

Stacks to be exposed are assembled from NTE layers of size up to 10$\times$20 cm$^2$ of a thickness of 200 $\mu$m on a glass base and 550 $\mu$m without it. If a beam is directed parallel to a layer plane, then tracks of all relativistic fragments remain long enough in one layer for the 3-dimensional reconstruction of their emission angles. The base provides ``stiffness'' of the tracks, and its absence allows for longer tracking, including transitions to adjacent layers. Factors for obtaining significant event statistics are the stack thickness and the total solid angle of detection. NTE contains in similar concentrations of atoms AgBr and CNO and 3 times more H ones. In terms of hydrogen density, the NTE material is close to the liquid hydrogen target. This feature makes it possible to compare under the same conditions break-ups of projectile nuclei, both as a result of diffraction or electromagnetic dissociation on a heavy target nucleus, and as a result of collisions with protons. 

Searching nuclear interactions in NTE without sampling (or the ``following the track'' method) provides a fairly uniform detection efficiency of all possible types of interactions and allows one to determine the mean free path for a certain type of interaction. This method is implemented in tracking the beam tracks of the nuclei under study from the point of entry into the NTE layer to the interaction or to the exit of the track from the layer. Time consuming, it provides the best viewing quality and consistency. Statistics of several hundreds of peripheral interactions with certain configurations of relativistic fragments is achievable with transverse scanning.

Relativistic fragments are concentrated in the cone $sin\theta_{fr}$ = $p_{fr}/p_{0}$, where $p_{fr}$ = 0.2 GeV/\textit{c} is the measure of the nucleon Fermi momentum in the projectile nucleus, and $p_0$ is its momentum per nucleon. The charges of relativistic fragments $Z_{fr}$ = 1 and 2 the most important ones in the unstable state problem are determined visually due to the apparent difference in ionization. The charges \mbox{$Z_{fr}$ $\geq$ 3} are determined from the density of discontinuities or the electron track density. The condition for selection of peripheral interactions is the preservation by relativistic fragments of the projectile nucleus charge $Z_{pr}$, that is, $Z_{pr}$ = $\Sigma Z_{fr}$.  These interactions are a few percent of the inelastic ones.

With a measuring base of 1 mm, the resolution for tracks of relativistic fragments is no worse than 10$^{–3}$ rad. The transverse momentum $P_{\mathrm{T}}$ of a fragment with a mass number $A_{fr}$ is defined as $P_{\mathrm{T}}$ $\approx$ $A_{fr}p_{0}sin\theta$ in the approximation of conservation of the velocity of the primary nucleus (or $p_0$).  In the fragmentation of nuclei constituting NTE tracks of \textit{b}-particles ($\alpha$-particles and protons with energy below 26 MeV), \textit{g}-particles (protons with energy above 26 MeV), and also \textit{s}-particles (produced mesons) can be observed. The most peripheral interactions, called coherent dissociation or ``white'' stars, are not accompanied by fragmentation of target nuclei and the production of mesons. Photos and videos of characteristic interactions are available \cite{15,28}.

The mass numbers $A_{fr}$ of the relativistic fragments H and He are defined as $A_{fr}$ = $P_{fr}\beta_{fr}c/(P_{0}\beta_{0}c)$, where \textit{P} is the total momentum, and $\beta$\textit{c} is the velocity. The $P\beta c$ value is extracted from the average Coulomb scattering angle in NTE estimated from the track displacements at 2–5 cm lengths. To achieve the required accuracy it is necessary to measure the displacements in at least 100 points. The total momentum can be measured up to 2 to 50 GeV/\textit{c}. Energy of 10 \textit{A} GeV is the limit for identifying He fragments.

The mass number assignment to H and He fragment tracks is possible by total momentum values derived from the average angle of Coulomb scattering. The use of this laborious method is justified in special cases for a limited number of tracks. In the case of dissociation of stable nuclei, it is often sufficient to assume the correspondence of He - $^4$He and H - $^1$H since the established $^3$He and $^2$H contributions do not exceed 20\%. This simplification is especially true in extremely narrow $^{8}$Be and $^{9}$B decays \cite{6}.

The invariant mass of a system of relativistic fragments is defined as the sum of all products of 4-momenta $P_{i,k}$ fragments $M^{*2}$ = $\Sigma(P_i\cdot P_k)$. Subtracting the mass of the initial nucleus or the sum of masses of fragments $Q = M^* - M$ is a matter of convenience. The components $P_{i,k}$ are determined in the approximation of conservation of the initial momentum per nucleon by fragments. Reconstruction by the invariant mass of decays of relativistic unstable nuclei $^{8}$Be and $^{9}$B, mastered in the BECQUEREL experiment, confirmed the validity of this approximation. 

The most accurate measurements of the angles are provided with KSM-1 microscopes (Carl Zeiss, Jena) when using the coordinate method. Measurements are carried out in a Cartesian coordinate system. The NTE layer unfolds in such a way that the direction of the analyzed primary track coincides with the microscope stage axis OX with a deviation not worse 0.1–0.2 $\mu$m per 1 mm of track length. Then the axis OX coincides with the primary track projection on the layer plane, and the axis OY on it is perpendicular to the primary track. The axis OZ is perpendicular to the layer plane. The measurements along OX and OY are made with horizontal micro-screws, and along the OZ, the depth of field micro-screw is used. Three coordinates are measured on the primary and secondary tracks at lengths from 1 to 4 mm in increments of 100 $\mu$m, according to a linear approximation of which the planar and dip angles are calculated. Details and illustrations of measurements on the plane of the layer and its depth have recently been published \cite{18}.

\section{Relativistic decays of $^{8}$Be}
\label{sec:3}

In the fragmentation of relativistic nuclei in NTE intense tracks are observed often that branch into pairs of He tracks with minimal opening angles which are attributed $^{8}$Be decays. Obviously, such a definition is inconvenient when comparing data obtained at different values of the primary energy. Universal $^{8}$Be identification by the 2$\alpha$-pair invariant mass $Q_{2\alpha}$ is the first ``key'' to the problem of the unstable nuclear states.

\begin{table}
	\caption{}
	\label{tab:1}       
	\begin{tabular}{ccc}
		\hline\noalign{\smallskip}
		$Q_{2\alpha}$, MeV 
		&\begin{tabular}{cc}
			$N_{2\alpha}$ (\%) \\
			$n_g$ + $n_b$ + $n_s$ = 0 
		\end{tabular}  
		& \begin{tabular}{cc}
			$N_{2\alpha}$ (\%) \\
			$n_g$ + $n_b$ + $n_s$ $>$ 0 
		\end{tabular}  \\
		\noalign{\smallskip}\hline\noalign{\smallskip}
		$Q_{2\alpha}$ $\leq$ 0.2  & 81 (41 $\pm$ 6) & 103 (34 $\pm$ 4) \\
		0.2 $<$ $Q_{2\alpha}$  $\leq$ 1 & 48 (24 $\pm$ 6) &	40 (13 $\pm$ 2) \\
		1 $<$ $Q_{2\alpha}$  $\leq$ 5	& 55 (27 $\pm$ 4) &	108 (36 $\pm$ 4) \\
		$Q_{2\alpha}$ $>$ 5 & 14 (7 $\pm$ 2) &	51 (17 $\pm$ 3) \\
		\noalign{\smallskip}\hline
	\end{tabular}
\end{table}

\begin{figure}
	\resizebox{0.7\textwidth}{!}{%
		\includegraphics{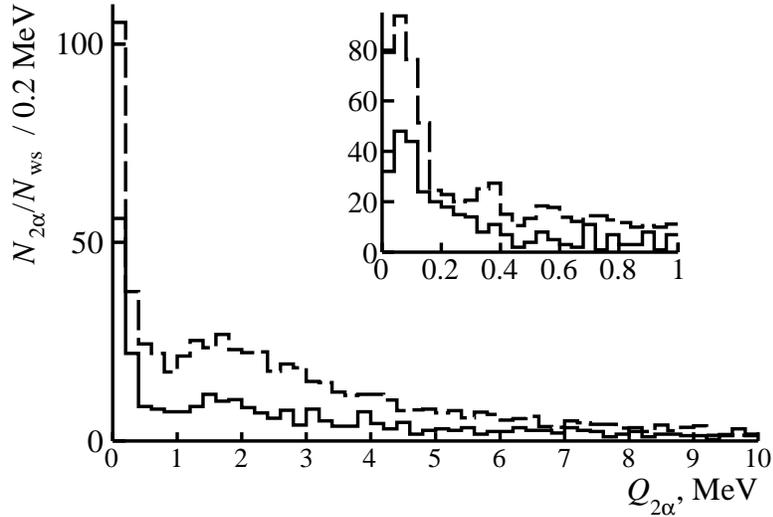}
	}
	\caption{Distribution of the number of 2$\alpha$-pairs $N_{2\alpha}$ over invariant mass $Q_{2\alpha}$ in coherent dissociation $^{12}$C $\to$ 3$\alpha$ (solid line) and $^{16}$O $\to$ 4$\alpha$ (dashed line) at 3.65 $A$ GeV; the inset, enlarged part $Q_{2\alpha}$ $<$ 1 MeV (step 40 keV); histograms are normalized to the number of ``white'' stars $N_\mathrm{ws}$.}
	\label{fig:1}       
\end{figure}

The distribution $Q_{2\alpha}$ is shown in Fig. \ref{fig:1} for the coherent dissociation $^{12}$C $\to$ 3$\alpha$ and \mbox{$^{16}$O $\to$ 4$\alpha$} at 3.65 \textit{A} GeV. In the $^{12}$C case, measurements of polar and azimuthal angles of $\alpha$-particles in 316 ``white'' stars made in the 90s by the groups of G. M. Chernov (Tashkent) \cite{29} and A. Sh. Gaitinov (Alma-Ata) and recently supplemented by the FIAN and JINR groups are used. In the $^{16}$O case, similar data is available for 641 ``white'' stars \cite{30}. For these events, Fig. \ref{fig:1} presents distributions of invariant mass in the region $Q_{2\alpha}$ $<$ 10 MeV of all 2$\alpha$-pair combinations $N_{2\alpha}$ normalized to the corresponding number of ``white'' stars $N_\mathrm{ws}$. In the insert these data are shown in the range $Q_{2\alpha}$ $<$ 0.5 MeV in an enlarged form. Although in both cases there are peaks corresponding to $^{8}$Be, however, due to the presence of ``tails'' caused by reflections of (3-4)$\alpha$ excitations the selection condition $Q_{2\alpha}$($^{8}$Be) is not sufficiently defined.

\begin{figure}
	\resizebox{0.7\textwidth}{!}{%
		\includegraphics{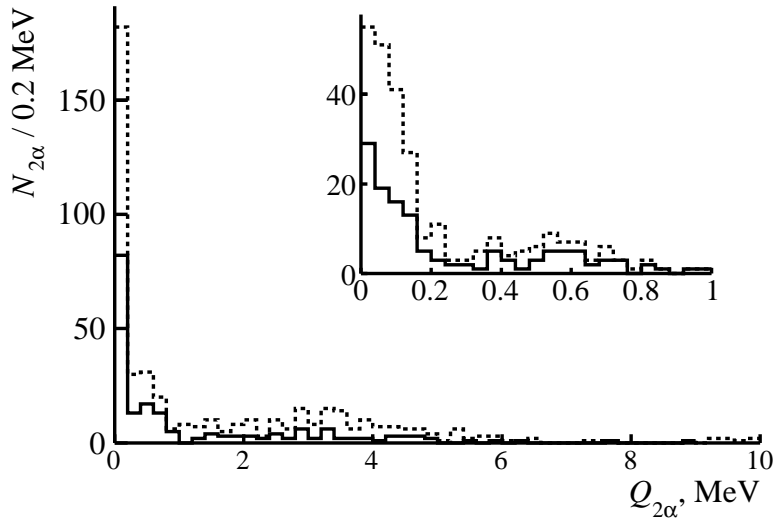}
	}
	\caption{Distribution of the number of 2$\alpha$-pairs of $N_{2\alpha}$ over the invariant mass $Q_{2\alpha}$ in 500 dissociation events $^9$Be $\to$ 2$\alpha$ (dots) at 1.2 $A$ GeV including 198 ``white'' stars (solid line); in inset, enlarged part $Q_{2\alpha}$ $<$ 1 MeV.}
	\label{fig:2}
\end{figure}

The angular measurements of $^{9}$Be $\to$ 2$\alpha$ at 1.2 \textit{A} GeV makes it possible to refine the selection condition of $^{8}$Be decays by the invariant mass $Q_{2\alpha}$ (review \cite{15}). The distribution over $Q_{2\alpha}$ of 500 2$\alpha$-pairs including 198 ``white'' ones presented in Fig. \ref{fig:2} indicates the limit $Q_{2\alpha}$($^{8}$Be) $<$ 0.2 MeV. There are two ``influxes'' centered on $Q_{2\alpha}$ equal to 0.6 and 3 MeV. The first reflects the $^{9}$Be excitation at 2.43 MeV \cite{10,31}, and the second one – the $^{8}$Be 2$^+$ state \cite{10}. The condition $Q_{2\alpha}$($^{8}$Be) takes into account the accepted approximation, the kinematic ellipse of $^{8}$Be decay, and the resolution of angular measurements. Its application allows us to determine the contribution of $^{8}$Be decays to the statistics of ``white'' stars equal to \mbox{45 $\pm$ 4\%} for \mbox{$^{12}$C $\to$ 3$\alpha$} and 62 $\pm$ 3\% for $^{16}$O $\to$ 4$\alpha$. A similar selection of $^{12}$C $\to$ 3$\alpha$ at 0.42 \textit{A} GeV gives 53 $\pm$ 11\% \cite{18}. The condition $Q_{2\alpha}$($^{8}$Be) $<$ 0.2 MeV coincides with those adopted in the electronic experiments \cite{24,25,26,31}.

Table 1 gives the distribution over the characteristic $Q_{2\alpha}$ intervals of the number of $\alpha$ pairs in ``white'' stars and stars containing additional tracks. The bulk of the statistics corresponds to dissociation through $^{8}$Be 0$^+$ and 2$^+$ states in a close ratio which is especially pronounced in the case of the sample with additional tracks definitely referring to neutron removal. This fact corresponds to the description of $^{9}$Be, in which the $^{8}$Be core is presented as an equal mixture of the 0$^+$ and 2$^+$ states \cite{32,33}. The question of whether the contribution of the 2.43 MeV excitation reflects the presence of the component $\alpha$ + $\alpha$ + \textit{n} or whether it is a reaction product requires theoretical consideration.

To estimate the boundaries of assumptions in the identification of $^{8}$Be associated with attributing the mass number \textit{A} = 4 and the corresponding momentum to He fragments, the fragmentation $^{9}$Be $\to$ 2He + \textit{n} was modeled in Geant4 \cite{34} in the framework of the QMD \cite{35} model for 10$^4$ $^{9}$Be nuclei of energy around 1.2 \textit{A} GeV ($\sigma$ = 100 MeV in the Gaussian distribution). 466 interactions $^{9}$Be $\to$ 2He + \textit{n} were obtained, including 59 pairs $^{3}$He + $^{4}$He and 4 -- $^{3}$He + $^{3}$He. The value of the average momentum per nucleon of $^{4}$He fragments is 1915 MeV/\textit{c} at RMS 32 MeV/\textit{c}. In the $Q_{2\alpha}$ distribution, approximately 2/3 of the $^{9}$Be fragmentation events correspond to the formation of $^{8}$Be 0$^+$. The average value of the relative momentum difference in the $^{4}$He pars at $Q_{2\alpha}$ ($^{8}$Be) is 0.8\%, and the contribution of pairs with the participation of $^{3}$He to $Q_{2\alpha}$ $\leq$ 0.2 MeV is less than 3\%. All these facts indicate the validity of the assumptions made and the sufficiency of the precision angular measurements. Moreover, modeling indicates that the inclusion of momenta to determine $Q_{2\alpha}$ would make sense if the accuracy of their measurements is of the order of tenths of a percent, while maintaining the same angular resolution. Indeed, according to the data of the hydrogen bubble chamber, there is an $^{8}$Be peak in the distribution over the opening angle {6}, and, therefore, in $Q_{2\alpha}$. The inclusion of momenta in the calculation of $Q_{2\alpha}$, the measurement accuracy of which is estimated at 1.5\% at a length of 40 cm of liquid hydrogen, leads to the ``scattering'' of the peak. Undoubtedly, the use of momentum analysis also leads to an additional deterioration in the angular resolution. This conclusion is worth to be taken into account when planning electronic versions of such studies.

\section{Relativistic decays of $^9$B}
\label{sec:4}

\begin{figure}
	\resizebox{0.7\textwidth}{!}{%
		\includegraphics{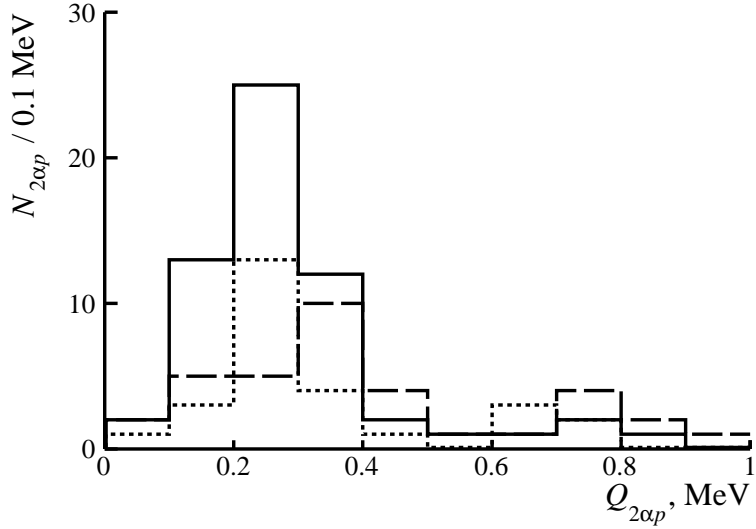}
	}
	\caption{Distribution of the number of 2$\alpha p$ triples $N_{2\alpha p}$ over invariant mass $Q_{2\alpha p}$ ($<$ 1 MeV) in events of coherent dissociation $^{10}$C $\to$ 2He + 2H (solid line) and dissociation $^{11}$C $\to$ 2He + 2H (dots) and $^{10}$B $\to$ 2He + H (dashed line).}
	\label{fig:3}       
\end{figure}

The next ``key'' in the unstable state studies is the $^{9}$B nucleus. When studying the coherent dissociation of the $^{10}$C isotope at 1.2 \textit{A} GeV, the 2He + 2H dissociation channel appeared as the leading one (review \cite{15}). The statistics of the 2He + 2H quartets in it amounted to 186 or 82\% of the observed ``white'' stars. The distribution over the invariant mass of 2$\alpha p$ triples $Q_{2\alpha p}$ presented in fig. \ref{fig:3} indicates the number of decays \textit{N}($^{9}$B) = 54 satisfying the condition $Q_{2\alpha p}$($^{9}$B) $<$ 0.5 MeV which is 30 $\pm$ 4\% of the events 2He + 2H. According to the condition $Q_{2\alpha}$($^{8}$Be) $<$ 0.2 MeV, $^{8}$Be decays are also identified in all these 2$\alpha p$ triples and only in them. This fact indicates the dominance of the decay sequence $^{9}$B $\to$ $^{8}$Be + p and $^{8}$Be $\to$ 2$\alpha$. The abundant formation of $^{9}$B nuclei in the dissociation of $^{10}$C indicates its important role as the structural basis of this isotope.

The confident identification of $^{8}$Be and $^{9}$B based on the $^{10}$C nucleonic composition allows one to turn to their contribution to the $^{10}$B and $^{11}$C dissociation. Angular measurements are performed in 318 events $^{10}$B $\to$ 2He + H at 1.0 \textit{A} GeV among which 20 decays $^{9}$B $\to$ $^{8}$Be + \textit{p} were identified that satisfy the condition $Q_{2\alpha p}$($^{9}$B) $<$ 0.5 MeV (Fig. \ref{fig:3}). Similarly, in 154 events $^{11}$C $\to$ 2He + 2H at 1.2 \textit{A} GeV \textit{N}($^{9}$B) = 22 (Fig. \ref{fig:3}) are found. Thus, in the dissociation of $^{10}$C, $^{10}$B and $^{11}$C, the universal condition $Q_{2\alpha p}$($^{9}$B) was established. In addition, when identifying $^{9}$B $\to$ $^{8}$Be + \textit{p} decays, the criterion $Q_{2\alpha}$($^{8}$Be) is confirmed under the purest conditions.

\section{Relativistic decays of the Hoyle state}
\label{sec:5}

\begin{figure}
	\resizebox{0.7\textwidth}{!}{%
		\includegraphics{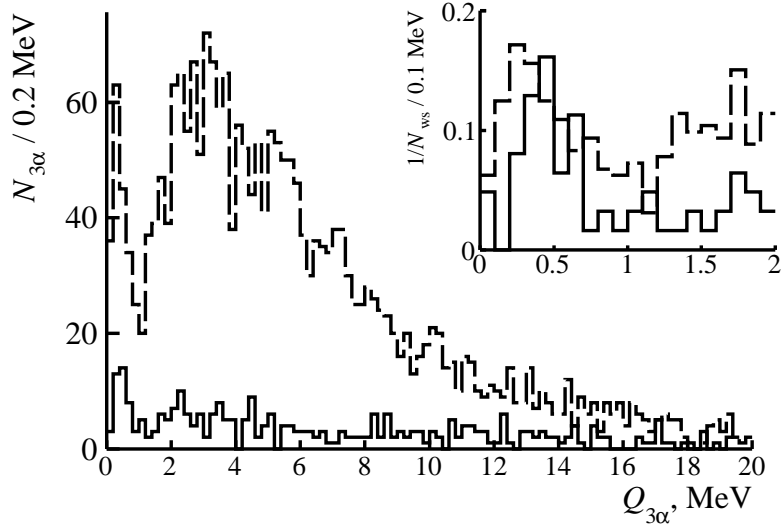}
	}
	\caption{Distribution of number of 3$\alpha$-triples $N_{3\alpha}$ over invariant mass $Q_{3\alpha}$ in 316 ``white'' stars $^{12}$C $\to$ 3$\alpha$ (solid) and 641 ``white'' stars $^{16}$O $\to$ 4$\alpha$ (dashed) at 3.65 $A$ GeV; in inset, enlarged part $Q_{3\alpha}$ $<$ 2 MeV normalized to numbers of ``white'' stars $N_\mathrm{ws}$.}
	\label{fig:4}
\end{figure}

Using the angular measurements of the ``white'' stars $^{12}$C $\to$ 3$\alpha$ and $^{16}$O $\to$ 4$\alpha$ the application of the invariant mass method can be easily extended to the identification of relativistic decays of the Hoyle state. In the latter case, HS decays can manifest themselves in the dissociation $^{16}$O $\to$ $^{16}$O$^{*}$ $\to$ $^{12}$C$^*$ ($\to$ 3$\alpha$) + $\alpha$. Both distributions over the invariant mass of 3$\alpha$-triples $Q_{3\alpha}$ presented in Fig. \ref{fig:4} show similarities. Their main parts in the region $Q_{3\alpha}$ $<$ 10 MeV, covering the $^{12}$C $\alpha$-particle excitations up to the nucleon separation threshold are described by the Rayleigh distribution with parameters $\sigma_{Q_{3\alpha}}$($^{12}$C) = 3.9 $\pm$ 0.4 MeV and $\sigma_{Q_{3\alpha}}$($^{16}$O) = 3.8 $\pm$ 0.2 MeV.

In both cases, distribution peaks are observed in the region $Q_{3\alpha}$ $<$ 0.7 MeV where the HS signal is expected. The statistics in the peaks minus the background is \mbox{$N_\mathrm{HS}$($^{12}$C) = 37} with an average value $\left\langle Q_{3\alpha} \right\rangle$ (RMS) = 417 $\pm$ 27 (165) keV and $N_\mathrm{HS}$ ($^{16}$O) = 139 with $\left\langle Q_{3\alpha} \right\rangle$ (RMS)  = 349 $\pm$ 14 (174). On this basis, the contribution of HS decay to the coherent dissociation of $^{12}$C $\to$ 3$\alpha$ is 11 $\pm$ 3\%, and in the case of $^{16}$O $\to$ 4$\alpha$, it is 22 $\pm$ 2\%. An increase in 3$\alpha$ combinations in $^{16}$O $\to$ 4$\alpha$ leads to a noticeable increase in the contribution of HS decays. At the same time, the ratio of the $^{8}$Be and HS yields shows an approximate constancy $N_\mathrm{HS}$($^{12}$C)/$N_\mathrm{^{8}Be}$($^{12}$C) = 0.26 $\pm$ 0.06 and $N_\mathrm{HS}$($^{16}$O)/$N_\mathrm{^{8}Be}$($^{16}$O) = 0.35 $\pm$ 0.04.

\begin{figure}
	\resizebox{0.7\textwidth}{!}{%
		\includegraphics{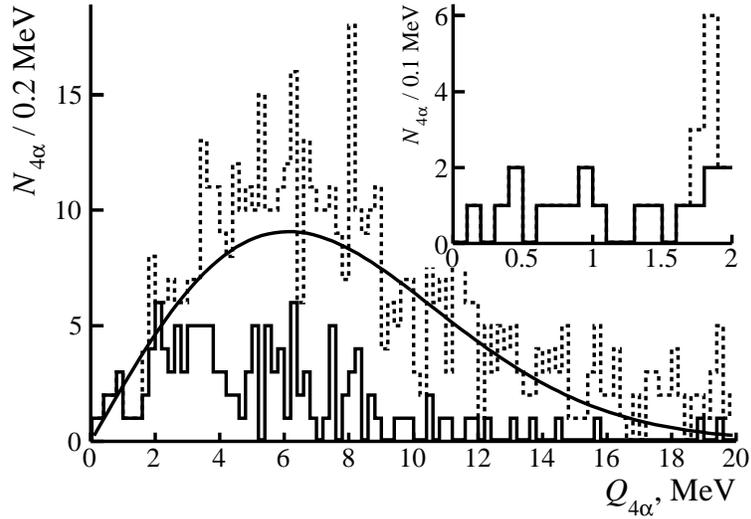}
	}
	\caption{Distributions over invariant mass $Q_{4\alpha}$ in 641 ``white'' stars $^{16}$O $\to$ 4$\alpha$ at 3.65 $A$ GeV of all 4$\alpha$-quartets (dots) and $\alpha$HS events (solid line); smooth line - Rayleigh distribution; the inset, enlarged part $Q_{3\alpha}$ $<$ 2 MeV.}
	\label{fig:5}
\end{figure}

\begin{figure}
	\resizebox{0.7\textwidth}{!}{%
		\includegraphics{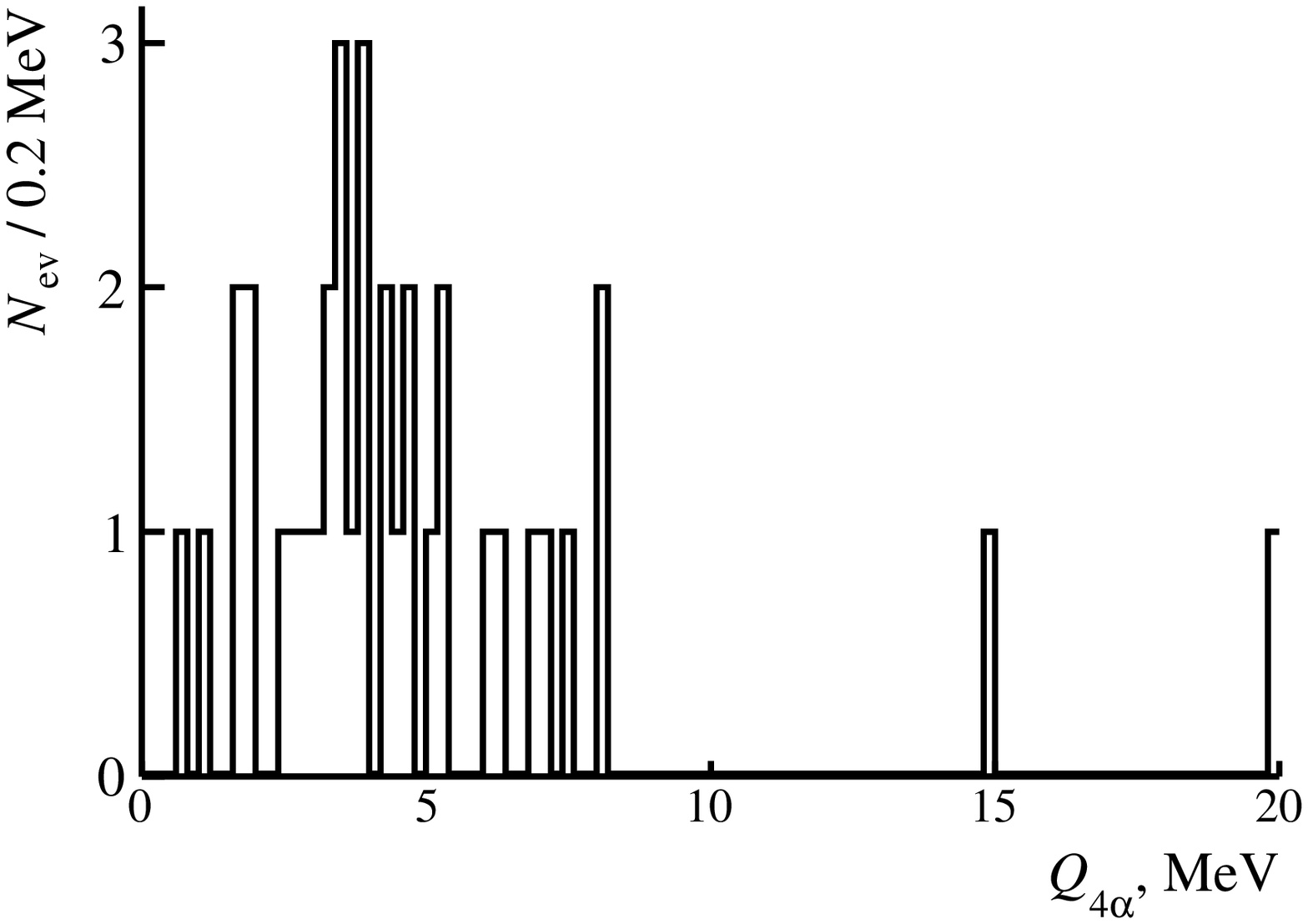}
	}
	\caption{Distribution of events $^{16}$O $\to$ 2$^8$Be over invariant mass $Q_{4\alpha}$.}
	\label{fig:6}
\end{figure}

There is a possibility of the HS emergence through the $\alpha$ decay of the 0$^+_6$ excitation of the $^{16}$O nucleus. The distribution of ``white'' $^{16}$O $\to$ 4$\alpha$ stars over the invariant mass of 4$\alpha$-quartets $Q_{4\alpha}$ presented in Fig. \ref{fig:5} in the main part is described by the Rayleigh distribution with the parameter $\sigma_{Q_{4\alpha}}$ = (6.1 $\pm$ 0.2) MeV. The condition for the presence of at least one 3$\alpha$-triple with $Q_{3\alpha}$(HS) $<$ 700 keV in a 4$\alpha$-event ($\alpha$HS) shifts the distribution over $Q_{4\alpha}$ to the low-energy side, and the parameter to $\sigma_{Q_{4\alpha}}$ = \mbox{4.5 $\pm$ 0.5} MeV (Fig. \ref{fig:5}). The enlarged view of the distribution over $Q_{4\alpha}$ presented in the inset in Fig. \ref{fig:5} indicates 9 events satisfying $Q_{4\alpha}$ $<$ 1 MeV and having an average value of $\left\langle Q_{4\alpha} \right\rangle $ (RMS) = 624 $\pm$ 84 (252) keV. Then, the contribution of the decays $^{16}$O(0$^+_6$) $\to$ $\alpha$ + HS is estimated to be 1.4 $\pm$ 0.5\% for normalization to $N_\mathrm{ws}$($^{16}$O) and 7 $\pm$ 2\% for normalization to $N_\mathrm{HS}$($^{16}$O). 

33 events $^{16}$O $\to$ 2$^{8}$Be are identified, which is \mbox{5 $\pm$ 1\%} of the ``white'' stars $^{16}$O $\to$ 4$\alpha$. Then, the statistics of coherent dissociation for the $^{16}$O $\to$ 2$^{8}$Be and $^{16}$O $\to$ $\alpha$HS channels has a ratio of 0.22 $\pm$ 0.02. The distribution over the invariant mass $Q_{4\alpha}$ of the events $^{16}$O $\to$ 2$^{8}$Be shown in Fig. \ref{fig:6} indicates two candidates $^{16}$O(0$^+_6$) $\to$ 2$^{8}$Be in the region of \mbox{$Q_{4\alpha}$ $<$ 1.0 MeV}. Thus, the estimate of the probability ratio of the channels $^{16}$O(0$^+_6$) $\to$ 2$^{8}$Be and $^{16}$O(0$^+_6$) $\to$ $\alpha$HS is 0.22 $\pm$ 0.17 which is too vague.

It can be concluded that although the direct dissociation dominates in the formation of HS a search for its 4$\alpha$ ``precursor'' is possible in the relativistic dissociation of nuclei. At the same time, increasing the statistics of events $^{16}$O $\to$ 4$\alpha$ in the traditional way can be considered exhausted. There remains the possibility of studying (3-4) $\alpha$-ensembles in the fragmentation of heavier nuclei.

\begin{figure*}
	\center{\includegraphics[width=0.75\linewidth]{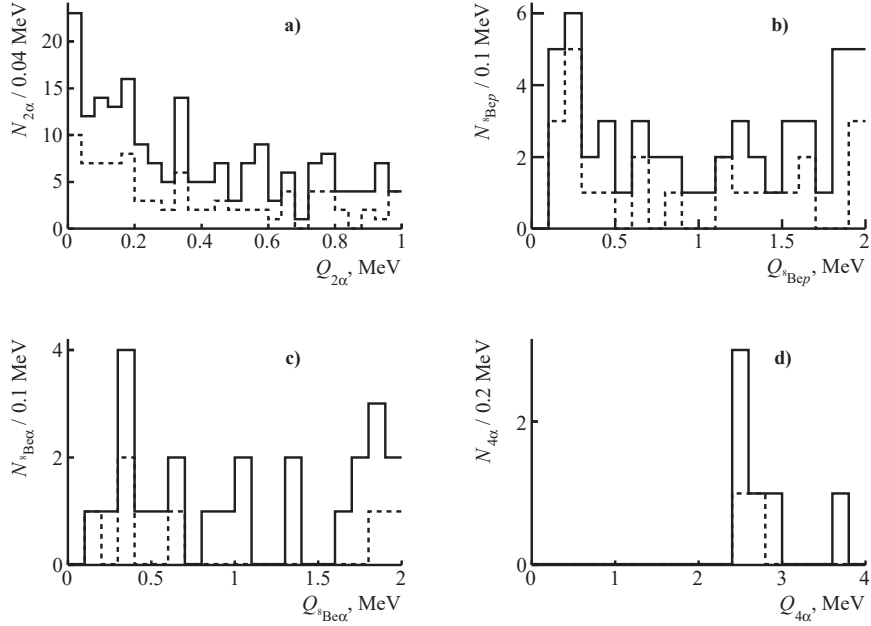}}
	\caption{Distributions of 2$\alpha$, 2$\alpha p$, 3$\alpha$ and 4$\alpha$ combinations from events of fragmentation of $^{28}$Si nuclei at 14.6 $A$ GeV over the invariant masses $Q_{2\alpha}$ (a), $Q_{2\alpha p}$ (b), $Q_{3\alpha}$ (c) and $Q_{4\alpha}$ (d) in the their small value regions. According to data for without sampling (points) and recent measurements in accelerated search (added by solid line).}
	\label{fig:7}       
\end{figure*}

\section{Search for unstable states in the fragmentation of Si and Au nuclei}
\label{sec:6}

\begin{figure*}
	\center{\includegraphics[width=0.75\linewidth]{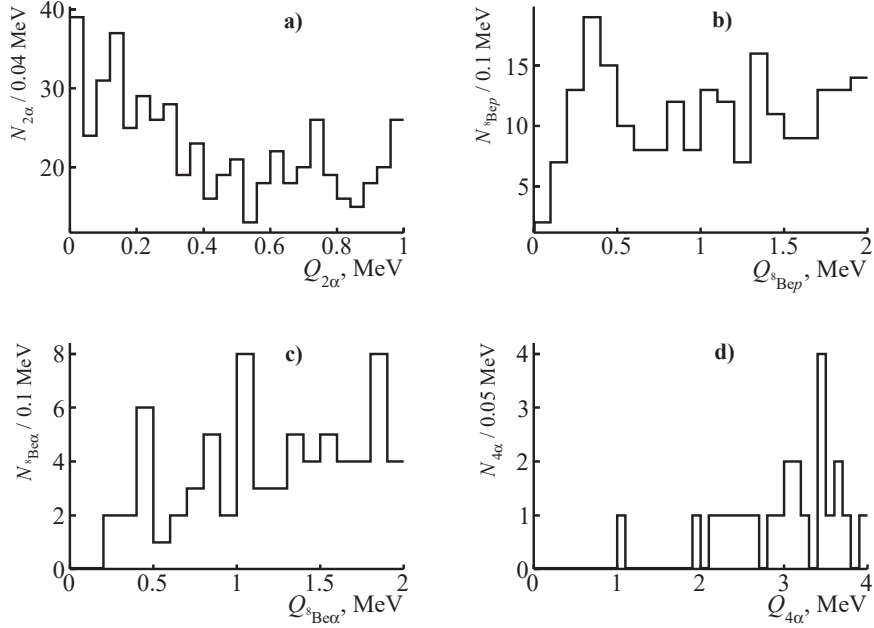}}
	\caption{Distributions of 2$\alpha$, 2$\alpha p$, 3$\alpha$ and 4$\alpha$ combinations from events of fragmentation of $^{197}$Au nuclei at 10.7 $A$ GeV over the invariant masses $Q_{2\alpha}$ (a), $Q_{2\alpha p}$ (b), $Q_{3\alpha}$ (c) and $Q_{4\alpha}$ (d) in the their small value regions.}
	\label{fig:8}       
\end{figure*}

EMU collaboration data are available on 1093 interactions of $^{28}$Si nuclei at 14.6 \textit{A} GeV \cite{36} and 1316 ones of $^{197}$Au at 10.7 $A$ GeV \cite{37} which contains measurements of the angles of emission of relativistic fragments. Then the search for events was conducted on the primary tracks without sampling. The number of events with the multiplicity of relativistic $\alpha$ particles $N_{\alpha}$ $>$ 2 in the Si case is 118, and Au is 843. Recently, the search for Si interactions $N_{\alpha}$ $>$ 2 has been resumed by scanning along the bands across the direction of entry of the primary tracks. The measurement cone is limited to 2$^{\circ}$. Thus, the analysis in the region of interest of small invariant masses is radically accelerated. In a relatively short time, 175 events $N_{\alpha}$ $>$ 2 were added to the Si statistics.The distributions over invariant masses $Q_{2\alpha}$, $Q_{2\alpha p}$, $Q_{3\alpha}$ and $Q_{4\alpha}$ in the small value regions obtained on the basis of these data are presented in Fig. \ref{fig:7} and \ref{fig:8}. According to the criteria described above, the numbers of decays $^8$Be ($N_\mathrm{^{8}Be}$), $^9$B ($N_\mathrm{^{9}B}$) and HS ($N_\mathrm{HS}$) are determined by them.

For the Si interactions, $N_\mathrm{^{8}Be}$ = 60 is obtained; $N_\mathrm{^{9}B}$ = 10 with $\left\langle  Q_{2\alpha p} \right\rangle  $ (RMS) = 273 $\pm$ 26 (103) keV; $N_\mathrm{HS}$ = 10 with $\left\langle  Q_{3\alpha} \right\rangle  $ (RMS) = 403 $\pm$ 48 (152) keV at $N_\alpha$ = 3(3), 4(3), 5(3) and 6(1). $N_\mathrm{HS}$/$N_\mathrm{^{8}Be}$ = 0.17 $\pm$ 0.05 at $N_{\alpha}$ $>$ 2. 4$\alpha$ quartets are absent up to $Q_{4\alpha}$ $<$ 2.5 MeV

A similar analysis of Au interactions yielded $N_\mathrm{^{8}Be}$ = 160; $N_\mathrm{^{9}B}$ = 40 with  $\left\langle Q_{2\alpha p} \right\rangle $ (RMS) = 328 $\pm$ 16 (116) keV; $N_\mathrm{HS}$ = 12 with $\left\langle Q_{3\alpha} \right\rangle $ (RMS) = 435 $\pm$ 29 (106) keV at $N_\alpha$ = 4(2), 6(2), 7(2), 8(2), 9(1), 11(1), 16(1). There is one 4$\alpha$ quartet including HS with $Q_{4\alpha}$ = 1 MeV at $N_\alpha$ = 16. In 11 events, $^{8}$Be pair formation is identified.

Under the assumption of a power-law dependence on the charge of the parent nucleus Z, which determines the production of $\alpha$ particles, the $^8$Be and $^9$B yields nuclei grows approximately as Z$^{0.8}$. Such a behavior is close to volume type dependence. The ratio of these yields is approximately the same. Statistics of 3$\alpha$ triples $N_\mathrm{HS}$ is small for estimates.

It can be concluded that in the both cases, $^8$Be and $^9$B decays are identified and indications of HS formation is obtained, and in the Au case, one candidate for 4$\alpha$ decay of the $^{16}$O(0$^+_6$) state was found. Due to the fact that the measurements were made without sampling, they allow one to plan searches for the unstable states. A set of statistics of events $N_\alpha$ $>$ 2 will be significantly accelerated during transverse scanning.

\section{Comments on longer-term objectives}
\label{sec:7}

The results obtained make it possible to assess the prospects of the presented approach in modern problems of nuclear physics. Among the most important of them is the verification of theoretical ideas about matter arising from the fusion of nucleons in clusters that do not have excited states up to the coupling threshold \cite{38}. These are the lightest $^{4}$He nuclei, as well as deuterons, tritons and $^{3}$He nuclei. The evolution of the composition of the lightest isotopes is predicted at a nuclear density less than normal and a temperature of several MeV. Passing through such a phase may be necessary on the way to the synthesis of heavy nuclei. A look at the dissociation of relativistic nuclei with time reversal indirectly indicates the feasibility of such a transition.

In the parent nucleus reference system, lightest fragment distributions over energy cover the temperature range 10$^{8}$-10$^{10}$ K corresponding to phases from the red giant to the supernova. In the dissociation of heavy nuclei, an unprecedented variety of coherent ensembles of the lightest nuclei and nucleons is available. The observations of unstable states presented here substantiate the possibility of studying cluster matter up to the lower limit of nuclear temperature and density. Identification of $^{1,2,3}$H and $^{3,4}$He isotopes by multiple scattering allows expanding the analysis of cluster states in the direction of the properties of the rarefied matter. The transverse momenta of fragments are determined from the emission angles, which makes it possible to isolate the temperature components. The practical feasibility of a detailed study of relativistic cluster jets can serve as a motivation for assessing the applicability of the relativistic approach to the problem of the existence of cold and rarefied nuclear matter.

In the relativistic dissociation of heavy nuclei, the formation of light fragments occurs with a greater ratio of charge to mass number than that of the primary nucleus, causing the appearance of associated neutrons that manifest themselves in secondary stars. The frequency of such ``neutron'' stars should increase with an increase in the number of lightest nuclei in the fragmentation cone. The average range of neutrons in NTE is about 32 cm. Reaching dozens, the multiplicity of neutrons in an event can be estimated by proportionally decreasing the average path to the formation of the ``neutron'' stars at paths of the order of several centimeters. The accuracy of determining the coordinates of their vertices makes it possible to restore the angles of neutron emission, and, therefore, the transverse momenta in the approximation of the conservation of the initial velocity. Thus, it is possible to study the effects of the neutron ``skin''. Estimation of the yield of neutrons, as well as deuterons and tritons binding neutrons, can be of applied value.

\begin{figure}
	\resizebox{0.7\textwidth}{!}{%
		\includegraphics{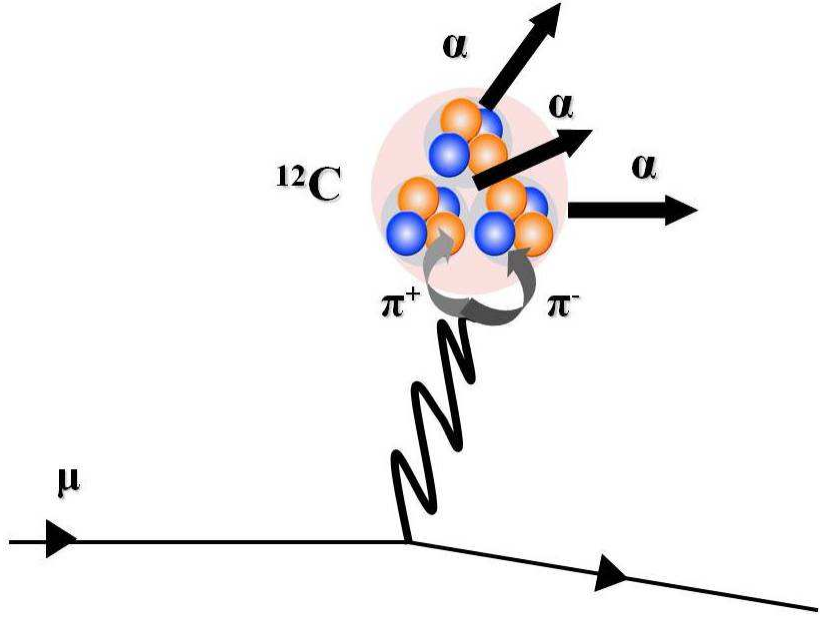}
	}
	\caption{Diagram of breakup of $^{12}$C nucleus into three $\alpha$-particles by relativistic muon.}
	\label{fig:9}       
\end{figure}

It remains unclear why the peripheral dissociation of the nuclei corresponds to a sufficiently large cross section and a wide distribution over the multiplicity of fragments. This phenomenon may be based on the transition of virtual photons exchanged between the beam and target nuclei into pairs of virtual mesons. A critical test can be the fragmentation of the NTE composition nuclei under the action of relativistic muons \cite{39,40,41}. The combination presented in fig. \ref{fig:9} provides long-range interaction with effective nuclear destruction and can be extended to peripheral interactions of relativistic nuclei. It was established that fragmentation of the target nuclei under the action of muons is most likely for the breakup $^{12}$C $\to$ 3$\alpha$. In these events, the $\alpha$ particle energy and emission angles are determined from the ranges making it possible to obtain distributions over the invariant mass, as well as over the total momentum of pairs and triples of $\alpha$ particles. It has been preliminary established that the distribution over the total transverse momentum of the $\alpha$-particle triples corresponds not to electromagnetic, but nuclear diffraction. Note that the 3$\alpha$ splitting cross section is important for geophysics, since it will allow testing the hypothesis of helium generation in the earth's crust by cosmic muons.

\section{Conclusions}
\label{sec:8}

Preserved and recently obtained data on interactions of light relativistic nuclei in a nuclear track emulsion allowed to establish the contribution in their dissociation of unstable nuclei $^{8}$Be and $^{9}$B and the Hoyle state as well as to assess the prospects of such research in relation to medium and heavy nuclei. These three states are uniformly identified by the invariant masses calculated from the measured angles of emission of He and H fragments under the assumption of conservation of the primary momentum per nucleon. 

The $^{8}$Be selection in dissociation of the isotopes $^{9}$Be, $^{10}$B, $^{10}$C, and $^{11}$C is determined by the restriction on the calculated value of the invariant mass of 2$\alpha$-pairs to 0.2 MeV, and the $^{9}$B 2$\alpha p$-triple mass up to 0.5 MeV. The certainty in the $^{8}$Be and $^{9}$B identification of became the basis for the search for decays from the Hoyle state in the dissociation $^{12}$C $\to$ 3$\alpha$. In the latter case, the 3$\alpha$ triple invariant mass is set to be limited to 0.7 MeV. The choice of these three conditions as ``cut-offs from above'' is sufficient because the decay energy values of these three states are noticeably lower than the nearest excitations with the same nucleon compositions, and the reflections of more complex excitations is small for these nuclei.

Being tested in the studies of the light nuclei, a similar selection is applicable to the dissociation of heavier nuclei to search for more complex states. In turn, the products of $\alpha$-partial or proton decay of these states could be the Hoyle state or $^{9}$B, and then $^{8}$Be. A possible decay variant is the occurrence of more than one state from this triple. In any case, the initial stage of searches should be the selection of events containing relativistic $^{8}$Be decays.

Dozens of $^{8}$Be and $^{9}$B decays are identified in the relativistic fragmentation cone of Si and Au nuclei. At the same time, the small number of 3$\alpha$ triples attributable to the decay of the Hoyle state which requires increasing statistics to the current $^{8}$Be equivalent. Then, the search for the excited state $^{16}$O(0$^+_6$) will become feasible. There are no fundamental problems along this path since there are a sufficient number of earlier exposed NTE layers, with transverse scanning of which the required $\alpha$ ensemble statistics is achievable. This whole complex of problems, united by questions of identification of unstable states, is in the focus of the application in the BECQUEREL experiment in the present time.

It is hoped that the rapid progress in image analysis will give a whole new dimension to the use of the NTE method in the study of nuclear structure in the relativistic approach. The solution of the tasks set requires investment in modern automated microscopes and the reconstruction of NTE technology at a modern level. At the same time, such a development will be based on the classical NTE method, the foundations of which were laid seven decades ago in cosmic ray physics.

\end{document}